# Synthesis of Nanocrystals of Long Persisting Phosphor by Modified Combustion Technique

**Harish Chander, D. Haranath[*], Virendra Shanker, Pooja Sharma**

*Luminescent Materials and Devices Group, Electronic Materials Division*
*National Physical Laboratory, Dr. K. S. Krishnan Road, New Delhi-110 012. (India)*

**Abstract**

Synthesis and characterization of Nanocrystalline long persistent $SrAl_2O_4:Eu^{2+}$, $Dy^{3+}$ phosphor via a modified combustion process has been presented in the paper. In this synthesis process, a mixture of respective metal nitrates, flux and combustible agent (urea/camphor) were thermally treated with slight modification at 400-600°C for about 5 minutes. It resulted in low-density voluminous mass in contrast to a solid lump by conventional solid-state method. The present work reports the changes made in the combustion process to achieve the homogenous incorporation of dopants and large-scale production of the nanophosphor in a short interval of time. The samples have been characterized for nanophase, structural and luminescent properties.

## 1. Introduction

Nanophase materials are being vigorously explored as most of the physical properties are size dependent and are markedly affected as the particle sizes tend to nanometer level. Phosphors are one of the materials that show promising behaviour when synthesized in nanophase. Many phosphors have been made in nanophase by employing different techniques [1-5]. Divalent europium activated alkaline earth aluminates are known to be efficient long persisting phosphors for their high quantum efficiency in the visible region. These are essentially interesting, as they do not involve any radioactive isotope [1]. Particularly $SrAl_2O_4:Eu^{2+}$, $Dy^{3+}$ phosphor exhibits very bright and long lasting phosphorescence (>50 h) with emission wavelength of 530-540 nm [2]. The synthesis of these materials for display applications with considerably good initial brightness and long afterglow has been a major goal of many research groups [3-5] all over the world both in industry and academia for well over a decade. Commonly followed method of preparation of these group of phosphors is solid state reaction technique, in which appropriate oxides/carbonates along with the dopants and fluxes are mixed and fired at temperatures around 1200-1500°C for a few hours. This treatment results in a highly sintered, dense and hard mass of phosphor, which is difficult to crush and grind. Combustion synthesis route gives a fluffy mass reducible to quite fine particles with almost no effort. Conditions prevailing during the processing should favour formation of fine particles in sub-micron region. Oxidizing atmosphere prevails in combustion process [6]. Incorporation of europium in bivalent state invariably requires reducing atmosphere [7]. With a view to develop a process for the instant synthesis of nanophase particles of the phosphor, we employed combustion route [8]. In the present work, an attempt to synthesize nanophase $SrAl_2O_4:Eu^{2+}$, $Dy^{3+}$ phosphor by combustion technique with slight modifications has been made. The samples have been characterized for nanophase, structural and luminescent properties.

---

[*] Corresponding author (D. HARANATH)
e-mail: haranath@mail.nplindia.ernet.in

## 2. Experimental details

### 2.1 Sample preparation

Nitrates of Sr, Al, Eu and Dy; boric acid and urea were taken as starting materials. Soluble salts of Sr and Al were taken in stoichiometric ratio of 1:1 by mole and the dopants Eu (1 mol%) and Dy (2 mol%) along with a suitable combustible agent like urea/camphor. The constituents were put in a pyrex container of 10-20 times larger volume and made to paste like consistency by adding DI water. The paste was mixed well for homogenization of the mass. The pyrex container was placed in a partially closed ceramic or quartz tube. Then the tube was placed in a furnace maintained at around 400-600°C. In about five minutes or by the time the set up attained the temperature of furnace, reaction started with bright yellow flame. The reaction continued only for a few seconds. As soon as the reaction was over, tube was taken out of the furnace and allowed to cool. Fluffy mass of the $SrAl_2O_4:Eu^{2+}$, $Dy^{3+}$ phosphor in nanocrystalline phase filling substantial part of pyrex container was obtained and characterized. Fig. 1 shows the experimental set up for combustion synthesis. The main purpose of valve is to release air into the firing tube at the time of sample removal as vacuum is developed during the cooling process.

## 2.2 Characterization

The phase purity and homogeneity of the combustion product was investigated by X-ray diffraction (XRD) technique. The XRD profiles were taken using a Brucker-AXS D8 advance diffractometer (with DIFFRAC plus software) using Cu Kα radiations. The particle size and morphological investigations of the phosphor prepared in the process were carried out with a scanning electron microscope (SEM, LEO 440 system). The spectral energy distributions were recorded using Perkin-Elmer Spectrophotometer (Model: LS-55) and a xenon flash lamp as the source of excitation. Initially, 365 nm was chosen as fixed wavelength for recording emission spectrum. With a filter at 515 nm, the emission spectrum was recorded in the range 400-900 nm. Using the wavelength of emission peak maximum, the excitation spectrum was recorded in the range 200-400 nm. The peak of maximum intensity was again chosen to record the emission spectrum. By repeating the procedure a couple of times a consistent set of excitation-emission spectra was obtained. Phosphorescence decay was recorded by collecting total light output using a photomultiplier tube (EMI 9658 B) and a DC microvoltmeter (Philips Model No. PP9004) after cutting off the excitation source. For decay studies, phosphors were excited by 200 W tungsten lamp for 5 minutes.

## 3. Results and discussion

Rare earth ions (Eu, Dy) doped strontium aluminate nanophosphors have been prepared by the combustion of respective nitrates along with urea at ~600°C. The aqueous mixture containing stoichiometric amount of redox mixture when heated rapidly at ~600°C boils and undergoes dehydration followed by decomposition generating combustible gases such as oxides of nitrogen, HNCO and $NH_3$. The volatile combustible gases ignite and burn with a flame and thus provide conditions for formation of phosphor lattice with dopants. The large amount of escaping gases dissipates heat and thereby prevents the material from sintering and thus provides conditions for formation of nanocrystalline phase. Also, as the gases escape they leave voluminous, foaming and crystalline fine powder occupying the entire volume of the firing container and have no chance of forming agglomerations unlike in the other conventional processes.

Fig. 2(a) and 2(b) show SEM micrographs of $SrAl_2O_4:Eu^{2+},Dy^{3+}$ phosphor prepared by solid-state reaction technique and combustion process respectively. Particle size of the phosphor obtained by combustion process is in sub-micron range and is very well crystalline as against molten glassy phase observed in case of solid-state reaction route. In case of combustion synthesis, instantaneous and in-situ very high temperature, combined with release of large volume of volatiles from liquid mixture is likely to result in production of nanoparticles in a fluffy form. Uniform nano-crystalline particles of SRAC can be observed in Fig. 2(b). Glassy nature for SRA is attributed to high temperature treatment (> 1400°C) for longer duration (> 5 h) in presence of flux, boric acid. For the SRA samples, irregularly rectangular grains with sizes of several tens of micrometers can be observed in Fig. 2(a). The grain boundaries with complete melting morphology can also be recognized.

In order to determine the crystal structure and to establish chemical nature of the combustible product, X-ray diffraction (XRD) study was carried out. The XRD patterns of the phosphors were taken using Cu Kα radiation at 35 kV tube voltage and 11 mA tube current. $SrAl_2O_4$ crystal is a monoclinic lattice. The XRD patterns of $SrAl_2O_4:Eu^{2+}$ prepared using solid state reaction method (SRA) and combustion synthesis (SRAC) are shown in Fig. 3(a) and 3(b) respectively. The peaks of (011), (-211), (220), (211) and (031) planes, which characterize $SrAl_2O_4$ crystal [9] were obtained in both the cases of Fig. 3(a) and 3(b). This confirms that the major phase present in the phosphor prepared by combustion method is $SrAl_2O_4$. However, some additional peaks in SRAC sample indicate partial presence of hexagonal phase of $SrO-Al_2O_3$ system [10].

The spectral energy distributions (SED) of the phosphors prepared by solid-state reaction and combustion techniques are shown in Fig. 4(a) and (b) respectively. Both SRA and SRAC samples show strong absorption in near UV region. It is clear from the figure that excitation peak maximum for SRAC sample shifts slightly towards shorter wavelength (from 343 to 337 nm) as compared to SRA sample. This shift may be due to the quantum size effect (QSE) of the phosphor nanoparticles. The size reduction should have widened the bandgap and hence, the SRAC samples need higher energy radiations to excite [11]. The emission spectra indicate the brightest luminescence at 529 nm for both SRA and SRAC samples. This emission peak maximum is the characteristic electronic transition of $Eu^{2+}$ ions between its $4f^6 5d^1 \rightarrow 4f^7$ levels [12]. Interestingly, a broad emission band peaking at around 667 nm has also been observed for SRAC samples. In general, the emission radiation of the phosphor is governed by either crystal field environment of luminescing ions or by the degree of covalence (coordination number) of these ions with the surrounding oxygen ions [13]. The exact explanation for the anomalous nature of the broad red emission is still not known. However, a detailed study is under progress to analyze this effect.

Fig. 5 shows comparative decay behavior of the phosphor samples prepared by two techniques. The rate of decay is comparatively faster in case of combustion samples to those observed in the sample made by solid-state reaction. Essentially, different decay rates for the SRA and SRAC samples shown in Fig. 5 indicate the presence of various types of traps with different depths in the samples. Flux, which is responsible for large sized particles due to crystallization, seems to add on to decay time for SRA. The observed enhancement of afterglow intensity and lengthening of decay time in SRA sample can be attributed to increase in number of defects due to reducing atmosphere and a higher concentration of $Eu^{2+}$ generated in the process [14]. Reduction in the particle size of the SRAC phosphor to sub-micron level is probably responsible for reduction in decay times [15]. Furthermore, the observed discrepancy of decay time of SRA and SRAC is probably related to $Eu^{2+}$ luminescent centres distributed in miscellaneous strontium borate or aluminate phases that may exhibit different strength of crystal field or different $Eu^{2+}$ coordination environments [13].

Using Lorentzian fit, the decay curve of SRAC sample was simulated by the following equation to understand the long persistent behavior of the phosphor [16].

$$I = I_o + A_1 \exp(-t)/\tau_1 + A_2 \exp(-t)/\tau_2 \qquad \ldots\ldots\ldots\ldots \quad (1)$$

Where I represent the phosphorescence intensity, $I_o$, $A_1$ and $A_2$ are the constants, t is the time, and $\tau_1$, $\tau_2$ are the decay constants of the phosphor. Fig. 6 represents its simulated curve and inset represents the list of parameters generated from it. The dots are the experimental points while the continuous line is the simulated curve produced from second order fitting. It is clear from the curve that decay data fits the equation very well. It can also be presumed from the above equation that there are two kinds of trapping levels with different depths present in the SRAC phosphor. Significant and varied values of $\tau_1$ and $\tau_2$ clearly indicate the concentration of shallow and deeper traps respectively. The higher initial intensity of the SRAC phosphor can be attributed to the presence of sufficient number of shallow traps while longer decay times to the deeper trap density.

However, a relevant rationalization regarding the enhanced phosphorescent decay for SRA compared to those for SRAC samples require further investigations such as thermoluminescence, photoconductivity and defect analysis, which are under progress.

## 4. Conclusions

The combustion method for the preparation of rare-earth doped nanocrystalline $SrAl_2O_4$ phosphor satisfies all the essential requirements of a long persistence phosphor. The formation of homogeneous single phase of monoclinic $SrAl_2O_4$ is confirmed by XRD analysis. The single step process to produce nanocrystalline long persistence phosphor in fluffy and voluminous form without agglomeration, which can be easily converted to fine and uniform powder, is an added advantage of this process. Further optimization of the process parameters can easily overrule the slight deviations observed in the decay times. We claim a process that can be easily adopted for producing phosphors in nanophase and facilitate efficient incorporation europium ions in 2+ states in any other commercially important host lattice also.

**Figure Captions**

Fig. 1: Experimental set up for combustion synthesis.

Fig. 2: SEM micrographs of (a) SRA and (b) SRAC phosphors.

Fig. 3: XRD patterns of the phosphor samples prepared by (a) solid-state reaction (SRA) and (b) combustion process (SRAC)

Fig. 4: Excitation (dotted line) and emission (solid line) spectra of (a) SRA and (b) SRAC phosphors.

Fig 5: Decay patterns of SRA and SRAC phosphors

Fig 6: Simulated decay pattern of SRAC phosphor. Inset represents the list of parameters generated from it.

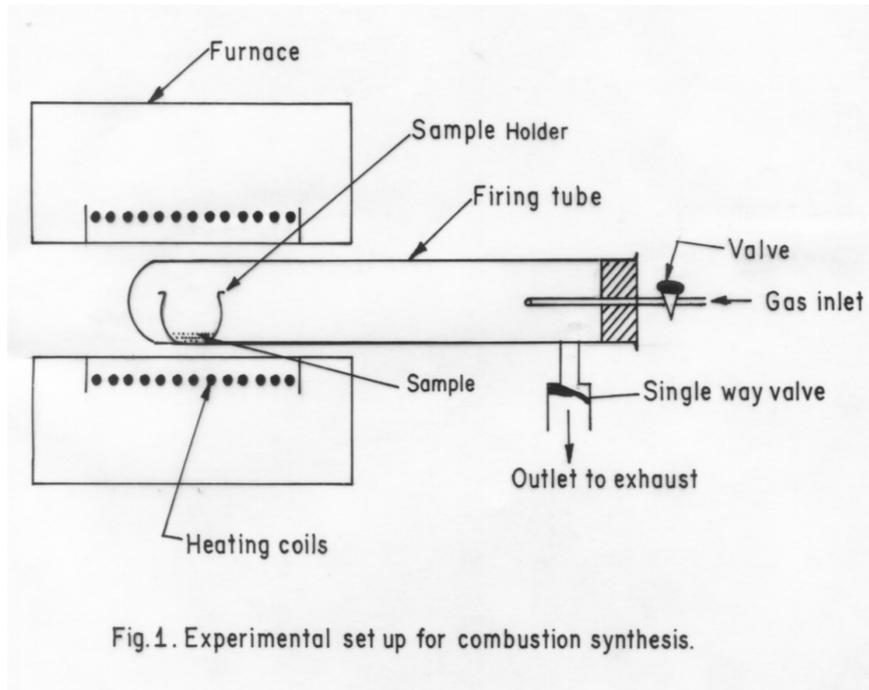

Fig. 1. Experimental set up for combustion synthesis.

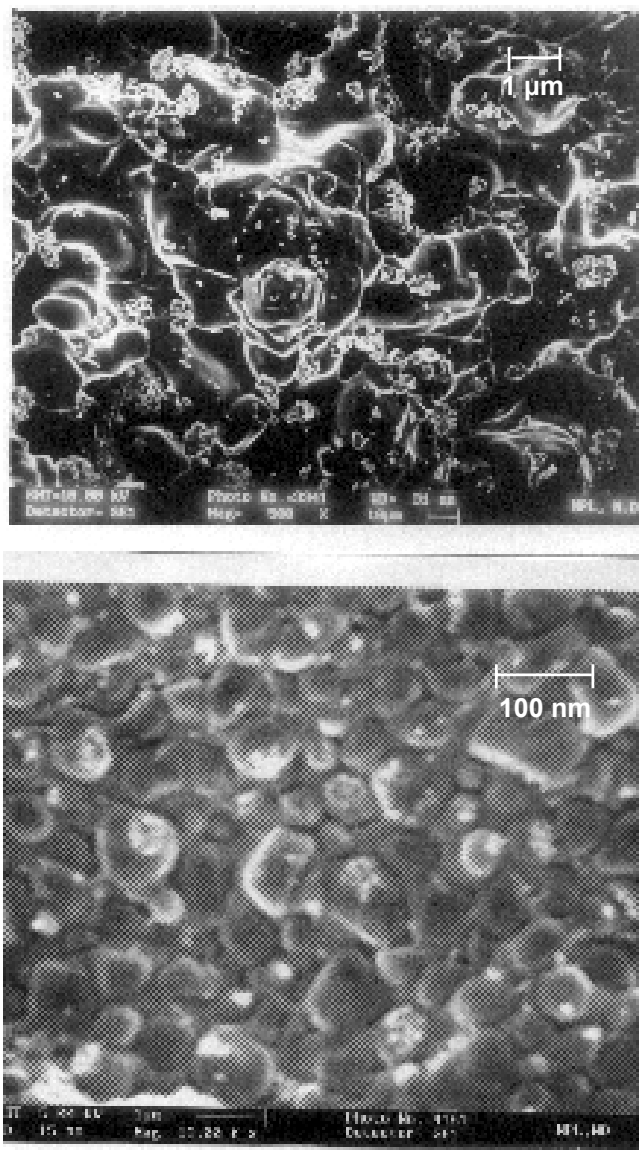

**Fig. 2 : SEM micrographs of (a) SRA and (b) SRAC phosphors.**

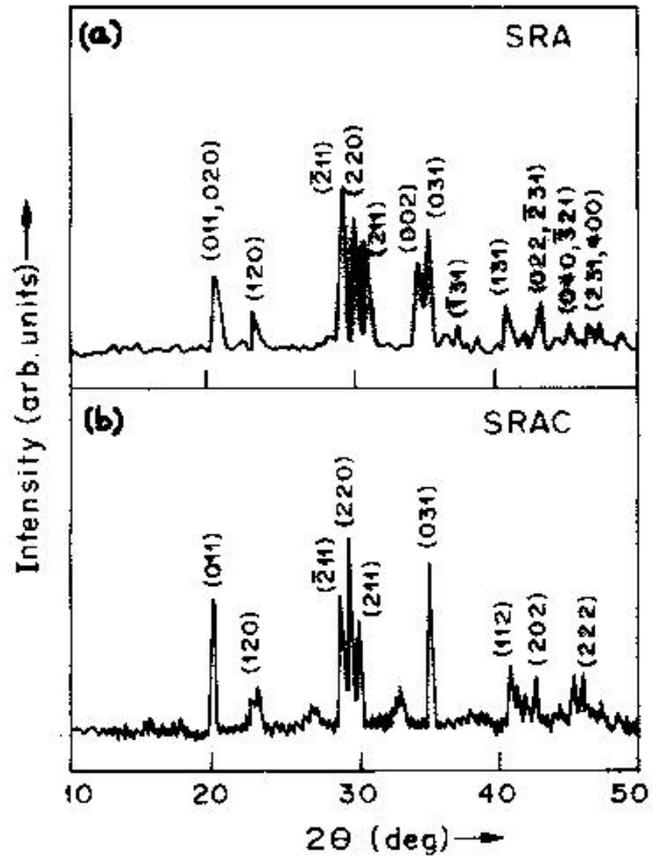

**Fig. 3 :** XRD patterns of the phosphor samples prepared by (a) solid-state reaction (SRA) and (b) combustion process (SRAC)

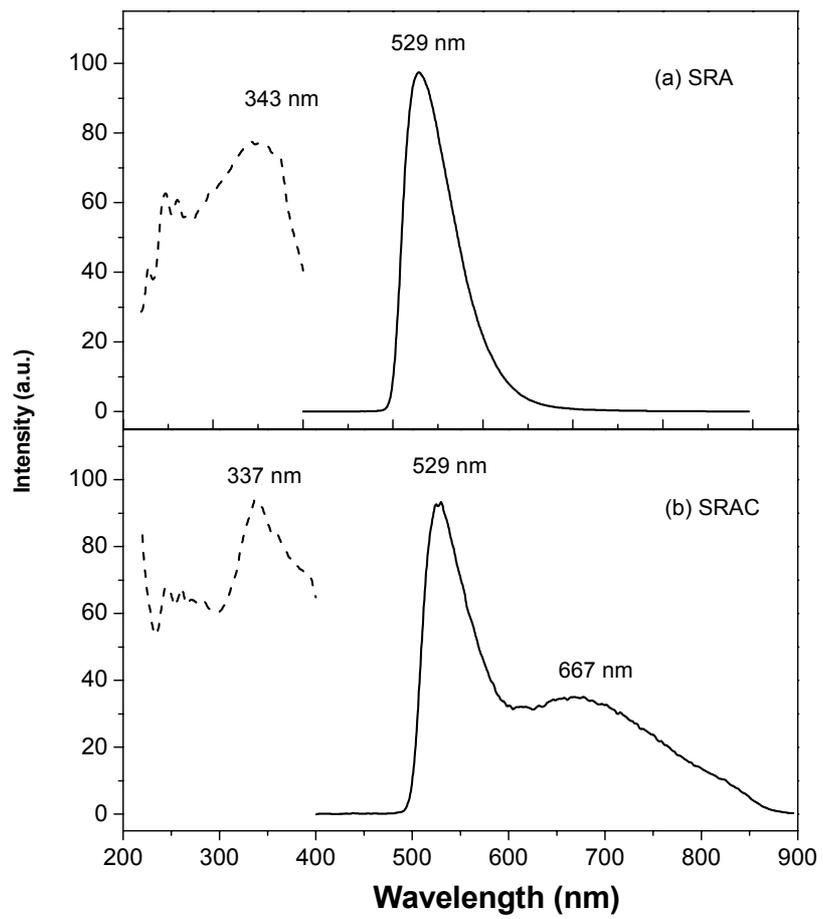

Fig. 4: Exciatation (dotted line) and emission (solid line) spectra of (a) SRA and (b) SRAC samples

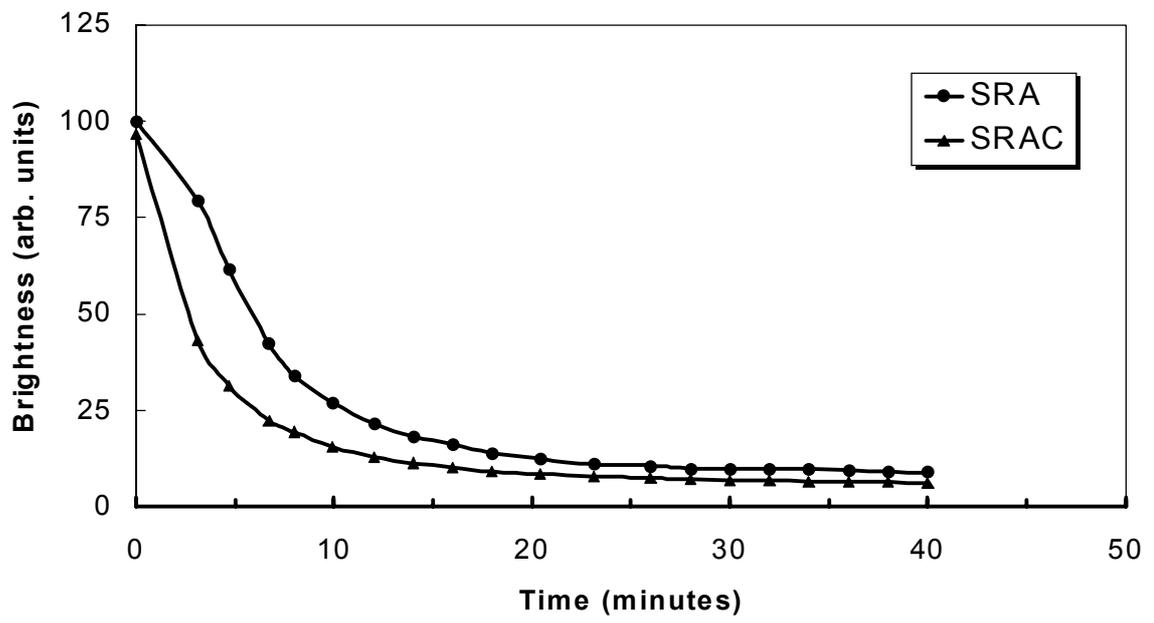

**Fig. 5: Decay patterns of SRA and SRAC phosphors**

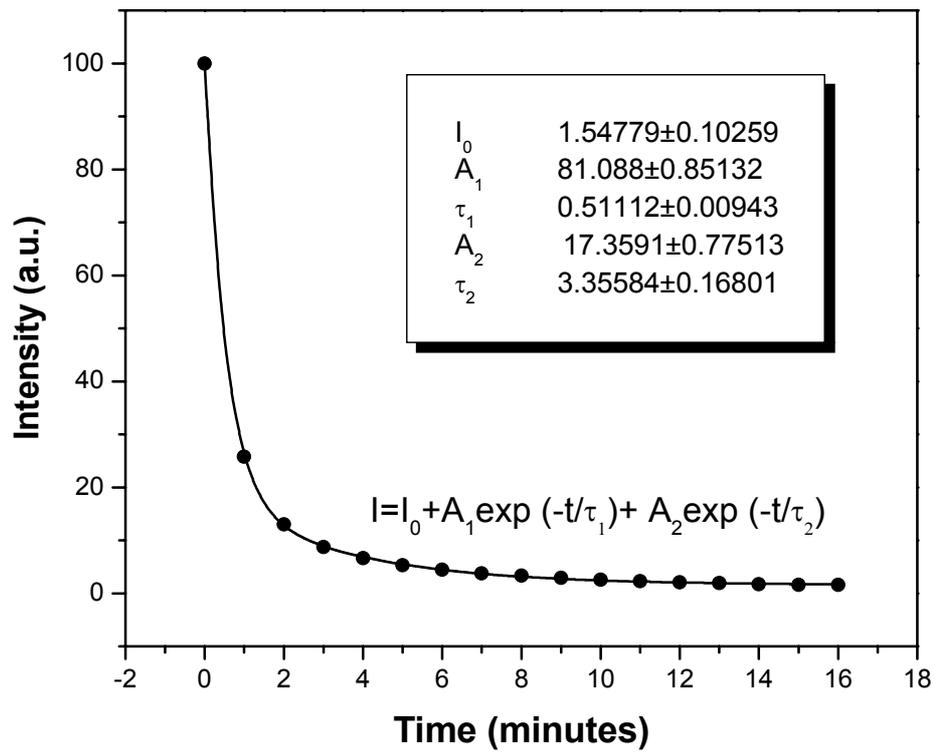

**Fig. 6: Simulation of decay behaviour of SRAC phosphor, dots representing the experimental values**